\documentclass[english,preprint]{ipsj}

\usepackage{graphicx}
\usepackage{latexsym}

\def\Underline{\setbox0\hbox\bgroup\let\\\endUnderline}
\def\endUnderline{\vphantom{y}\egroup\smash{\underline{\box0}}\\}
\def\|{\verb|}

\usepackage{amsmath}

\usepackage{listings}
\usepackage{xcolor}

\usepackage[varg]{txfonts}
\makeatletter%
\input{ot1txtt.fd}
\makeatother%

\usepackage{comment}
\usepackage{listings}
\usepackage{multirow} 
\usepackage[hang,small,bf]{caption}
\usepackage{subcaption}
\usepackage[hyphens]{url}

\title{PRA-TLS: Attestation of a Client Application for TEE}

\affiliate{Ocha}{Ochanomizu University}
\affiliate{ROIS}{Research Organization of Information and Systems Joint Support Center for Data Science Research, Center for R\&D on Secure Computer Systems}
\affiliate{Otsuma}{Otsuma Women's University}
\affiliate{NII}{National Institute of Informatics}

\author{Reina Sasaki}{Ocha,ROIS}[reina@ogl.is.ocha.ac.jp]
\author{Yutaka Ishikawa}{ROIS,Otsuma}[ishikawa.yutaka@rois.ac.jp]
\author{Atsuko Takefusa}{ROIS,NII}[takefusa@nii.ac.jp]
\author{Masato Oguchi}{Ocha,ROIS}[oguchi@is.ocha.ac.jp]

\begin{abstract}
Trusted Execution Environments (TEEs) are secure foundations for protecting sensitive information and executing computations over confidential data.
Processing in an isolated execution environment (enclave) is invoked by an untrusted application in the Rich Execution Environment (REE).
Then the enclave returns the execution results to the untrusted application.
Even if the enclave has been attested, the boundary between the enclave and the untrusted application poses risks, such as tampering with input arguments or return values and manipulating the order in which the application invokes functions.
Therefore, it is necessary to establish the authenticity and integrity of not only the enclave itself but also the application and its execution environment.
In this study, we propose an attestation protocol, Portable Remote Attestation TLS (PRA-TLS), for a client application that invokes an enclave.
We introduce a daemon that acts as an attester.
PRA-TLS uses a trusted Attester Daemon to measure the state of the host environment and application code at runtime, providing these measurements as attestation evidence for verification by a remote Verifier.
This mechanism allows the enclave to proceed only when the software environment is in the expected state and the application code is verified as legitimate.
We define attack models and security requirements for the proposed protocol and formally evaluate its security using the Tamarin Prover.
Furthermore, we implement a prototype of PRA-TLS using Intel SGX and evaluate its performance.

\end{abstract}


\begin{document}
\maketitle

\section{Introduction}
\label{intro}

With the widespread adoption of cloud services, it has become common to manage and process data containing personal and corporate confidential information in cloud environments.
Meanwhile, as cyberattacks become increasingly complex and diverse, cloud environments face a significant security challenge: protecting data confidentiality and integrity.
Traditionally, data at rest and data in transit have been protected using cryptographic mechanisms such as storage encryption and Transport Layer Security (TLS) \cite{rfc8446}, \cite{sinetstream2}.
However, data generally needs to be decrypted and loaded into memory before processing. 
This makes it difficult to protect data in use from the OS and other privileged software in the execution environment \cite{ccwhitepaper}.
That's why Confidential Computing \cite{cc} has been proposed to protect data in use, using Trusted Execution Environments (TEEs) \cite{tee}.

A TEE provides an isolated execution environment through hardware, such as a CPU or a dedicated chip. 
By processing data within a TEE, the risk of users leaking standard
keys due to external attacks, such as malware, is mitigated.
This enables data and executing code to be protected from an untrusted Rich Execution Environment (REE), even if an OS vulnerability is exploited to gain elevated privileges.

TEE implementations can be broadly classified into process-based and VM-based approaches \cite{sok24}.
Examples of process-based TEEs include Arm TrustZone \cite{arm}, Intel SGX \cite{sgx}, and RISC-V Keystone \cite{keystone}.
In this paper, we use the term enclave, following Intel SGX, to refer to the isolated execution environment provided by a TEE.
These process-based TEEs provide interfaces between a client application and an enclave, allowing operations to be invoked across the enclave boundary and input and output data to be exchanged \cite{sgxsdk, open_enclave, graphene-sgx}.
However, this boundary poses an attack surface risk because arguments in the inputs and outputs between the application and the enclave are poorly designed or maliciously manipulated \cite{tale_of_two, coin}.

Attestation allows a verifier to determine whether the enclave and its execution environment satisfy the expected security properties.
However, attestation primarily targets the code and state within the enclave, whereas the client application that invokes the enclave is typically outside its scope.
Consequently, even if an attacker invokes a legitimate enclave binary from a malicious client application, the enclave's attestation result may still be considered valid.
One possible approach to addressing this problem is to verify the client application's signature.
However, it is difficult to guarantee that the verified execution corresponds to the process actually invoking the enclave, or to ensure the integrity of the client application.

Therefore, even if an enclave securely processes confidential data, a compromised client application may still expose that data.
Thus, in addition to verifying the enclave itself, the verifier must also check the authenticity and integrity of the client application and confirm that a legitimate client application invokes the enclave.

In this study, we propose \textbf{Portable Remote Attestation TLS (PRA-TLS) }, an attestation protocol for client applications that invoke enclaves in TEEs, with a particular focus on process-based TEEs.
PRA-TLS realize the authenticity and integrity of the client application and its execution environment not only at startup but also during runtime.
To generate the evidence required to attest to the client application, we introduce a trusted Attester Daemon.
PRA-TLS extends a remote attestation protocol for Intel SGX, RA-TLS \cite{ra-tls}, to include the attestation of the client application.
We then define the attack models and security requirements for PRA-TLS and formally evaluate the protocol's security in the Tamarin prover \cite{tamarin-prover}.
Finally, we extend RA-TLS, a remote attestation protocol for Intel SGX, to support client-application attestation and evaluate the extension.

The contributions of this paper are as follows:
\begin{itemize}
    \item We propose a remote attestation protocol, PRA-TLS, for a client application that uses an enclave.
    \item We formally verify PRA-TLS using the Tamarin Prover to demonstrate that the protocol satisfies the defined security requirements under the assumed attack model.
    \item We implement the proof-of-concept of PRA-TLS and evaluate its performance.
\end{itemize}




\section{Related Work}

\subsection{TPM}
A Trusted Platform Module (TPM) is a hardware root of trust for the platform.
TPM provides protected storage for cryptographic keys and measurements. 
It also performs cryptographic operations such as key generation, key management, and digital signing. 
Private keys protected by the TPM are designed to remain within the TPM and are not exposed outside its protected boundary.
A TPM is provisioned with an Endorsement Key (EK), an asymmetric key pair whose private key is protected by the TPM and whose public key can be associated with an endorsement credential issued by the TPM manufacturer. 
The EK and its associated credential are used to establish the TPM's provenance and authenticity.
An Attestation Key (AK) is a signing key that attests to the platform's state to an external verifier. The EK allows the verifier to confirm that the AK belongs to a genuine TPM.
During the boot process, components such as the BIOS/UEFI firmware and OS kernel are cryptographically measured, typically by computing their hash values before control is transferred to the next stage; these values are called measurements.
These measurements are extended into Platform Configuration Registers (PCRs) within the TPM through TPM commands issued during the boot process.
TPM 2.0 defines 24 PCR indices, numbered 0 to 23, and platform-specific specifications define the intended uses of these PCRs \cite{tpm2-ptp}.
During attestation, the TPM generates a signed Quote containing selected PCR values, thereby providing evidence of the platform's measured state.
TPM sealing can bind access to sensitive data to a policy based on specific PCR values, so that the data can be unsealed only when the platform matches the expected measured state.

\subsection{Attestation}
\label{attestation}
Attestation can be broadly categorized into local attestation and remote attestation.
Local attestation enables two distinct entities on the same platform to verify each other's trustworthiness.
Remote attestation allows a remote Verifier to assess evidence about a platform, such as its system state and software configuration. A Relying Party then uses the result to determine whether the platform is trustworthy \cite{rats}.
The Verifier verifies the authenticity of the evidence by validating its signature and assesses the platform's integrity by appraising the evidence's contents.
Attestation requires evidence that enables assessment of the integrity of the system being attested.
In Intel SGX, attestation evidence is generated through the Quoting Enclave, whereas in TPM-based attestation, it is generated by the TPM hardware.
A Quote is generated by the Quote Enclave and is signed attestation evidence that contains the identity and measurement of an application enclave.

\subsubsection{Remote Attestation with TLS}
There are some TEE-based remote attestation protocols that combine with the TLS protocol \cite{ra-tls, ra-tls2, ra-tls3, ra-tls4}.
RA-TLS \cite{ra-tls} is a technology that integrates Intel SGX Remote Attestation into the standard TLS protocol.
In conventional Remote Attestation, attestation execution and the establishment of a secure channel via key exchange are performed as independent processes, which complicates the protocol and may introduce implementation vulnerabilities.
RA-TLS resolves this issue by embedding the SGX report (Enclave measurements and signatures) into the extension field of a standard X.509 certificate.
During the TLS handshake, the client or server verifies the report in the peer's certificate, confirming that the communication partner is a legitimate Enclave, while simultaneously completing the exchange of TLS session keys.
This eliminates complex verification logic at the application layer and allows for the establishment of a communication channel with high confidentiality and integrity, compliant with standard protocols, between an Enclave isolated from the OS and administrators.

\subsubsection{Attestation for Application}
Similar to our approach, systems have been proposed that use TPMs and the Integrity Measurement Architecture (IMA) \cite{linux_ima} to measure application integrity at runtime.
Keylime \cite{keylime} is an open-source project under the Cloud Native Computing Foundation (CNCF) \cite{cncf} that provides TPM-based attestation and runtime integrity measurement for distributed computing environments.
Keylime consists of four main components: an Agent, two server-side components, the Verifier and the Registrar, and a command-line tool, the Tenant.
The Agent acts as the prover and runs on the untrusted machine being attested, where it collects IMA measurements and obtains TPM Quotes.
The Verifier is responsible for evaluating attestation evidence and issues a revocation notification when an Agent is determined to be in an untrusted state.
The Registrar handles the initial registration of Agents by obtaining the EK public key, EK certificate, and AK public key, and verifying that the AK is associated with the TPM represented by the EK.
The Tenant is a command-line tool for managing nodes participating in the attestation process.

Operational challenges and security vulnerabilities have been identified in continuous integrity verification based on Keylime and IMA.
Ruffin et al. \cite{ruffin} experimentally demonstrated that unexpected OS updates can cause attestation failures, leading to false positives.
They also identified five vulnerabilities that allow attackers to evade IMA-based detection, resulting in false negatives.
Although these issues are mitigated through dynamically generated policies, the verification process still relies on continuous online monitoring.

Although our approach similarly leverages the TPM and IMA, it differs from prior systems and studies in that a daemon serves as the Attester.
Furthermore, our approach specifically targets client applications that use TEEs and differs in that it attests not only the TEE itself but also the client application that invokes it and its execution environment.

\section{Motivation and Design}
\label{motivation_and_design}

In this chapter, we first describe the motivation for attesting client applications that invoke enclaves in Section \ref{motivation}.
We then present the design goals and scope of PRA-TLS in Section \ref{design_goals}.

\subsection{Motivation}
\label{motivation}

\noindent{\large$\bullet$}\quad
\textbf{Challenges for Client Applications Invoking Enclaves}

\noindent
Process-based TEEs typically adopt a threat model in which software including the OS kernel is considered untrusted.
For example, Intel SGX creates a protected memory region called an enclave, in which sensitive data and code can be loaded and securely processed.
Intel SGX provides ECALL and OCALL interfaces, which enable function calls between enclave code and the outside.
An ECALL invokes an enclave function from a client application via an enclave entry transition initiated by the EENTER instruction.
In contrast, an OCALL invokes an external function, such as a system call, from within the enclave via an enclave exit transition using the EEXIT instruction.
The ECALL/OCALL interfaces at the boundary between the enclave and the client application have been shown to be susceptible to attacks such as tampering with ECALL arguments or OCALL return values, and to deliberately manipulating the order of ECALL invocations \cite{tale_of_two, coin}.
Because data and control flow across the enclave boundary through an untrusted environment, the ECALL/OCALL interfaces become part of the attack surface when the client application is compromised.
To mitigate such threats, the verifier must establish the client application's authenticity and integrity.

\noindent{\large$\bullet$}\quad
\textbf{Authenticity and Integrity Assurance for TEE Client Applications}

\noindent
In process-based TEEs such as Intel SGX and Keystone, attestation targets the enclave's code and initial state, while the client application that uses the enclave is outside the scope of attestation.
Therefore, the enclave's attestation result alone cannot directly establish the authenticity and integrity of the client application actually using the enclave.
In other words, even if an attacker invokes a legitimate enclave binary from a tampering client application, the enclave's attestation result may still be considered valid.

One possible approach to verifying the authenticity and integrity of the client application is to verify the signature of its executable.
However, enclave programs may be provided as shared objects (.so files) in Intel SGX and can therefore be invoked by arbitrary applications.
Therefore, if the enclave verifies a client application's signature, the enclave must maintain trusted information to identify authorized signers or the application code, and provide a secure mechanism to bind the verified code to the process that invokes the enclave.
Furthermore, client applications may use shared libraries at runtime. This makes it difficult to identify and verify all code that may be loaded and executed in advance.
Consequently, even if attestation establishes the enclave's authenticity and integrity, it remains difficult to determine whether the enclave is processed by the intended legitimate application.

Therefore, process-based TEEs, which typically assume a threat model in which software, including the client application, is untrusted, face the following challenges.
\begin{itemize}
    \item Data crossing the boundary between the enclave and the client application may be subject to tampering or no-intent ordering attacks.
    \item It is difficult to establish the authenticity and integrity of the client application that invokes the enclave.
\end{itemize}

Therefore, it is necessary to consider a mechanism for verifying the authenticity and integrity of the client application that invokes the enclave.

\subsection{Design Goals}
\label{design_goals}

In this study, we propose PRA-TLS, which is an attestation protocol for verifying the authenticity and integrity of client applications that invoke enclaves.

To address the challenges described in Section \ref{motivation}, PRA-TLS aims to provide the following two security guarantees.
\begin{itemize}
    \item Assurance of the authenticity and integrity of the execution environment in which the client application runs.
    \item Assurance of the authenticity and integrity of the client application not only at startup but also during runtime.
\end{itemize}

PRA-TLS extends RA-TLS, an existing remote attestation protocol for enclaves, to attest not only the enclave but also the client application.
PRA-TLS's system introduces an Attester Daemon that generates attestation evidence for the client application.
TPM is used as the hardware root of trust.
At boot time, UEFI Secure Boot verifies the signatures of boot components, including the bootloader and Linux kernel, to ensure that only authorized binaries are executed along the boot chain.
The Linux IMA is used to measure software components at runtime, including the Attester Daemon.
The Attester Daemon obtains a hash of the application code not only at system deployment but also at runtime when the client application invokes the enclave.
PRA-TLS is designed as a generic client application attestation protocol for process-based TEEs.


\section{System Model}
\begin{figure}[t]
\centerline{\includegraphics[width=\linewidth]{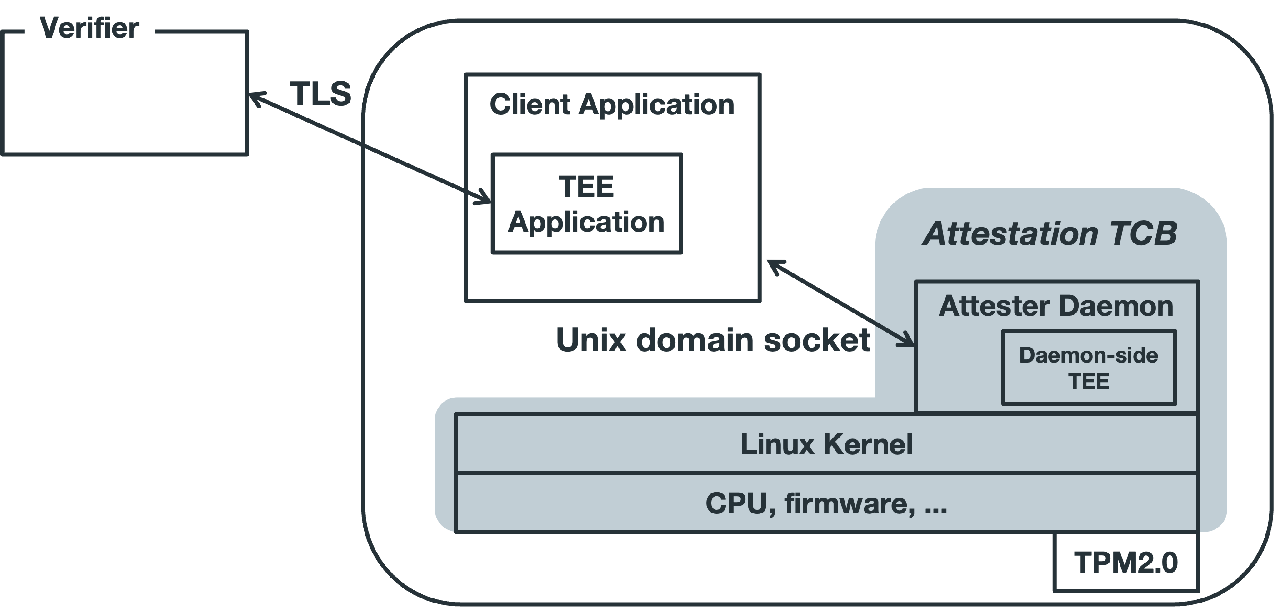}}
 \caption{The Architecture of the PRA-TLS system.}
 \label{system}
\end{figure}

\figref{system} illustrates the architecture of the PRA-TLS system.
The PRA-TLS system consists of a remote Verifier; the software stack from the CPU to the Linux kernel; a Client Application running on this stack; a TEE Application invoked by the Client Application; an Attester Daemon; a Daemon-side TEE invoked by the Attester Daemon; and a TPM 2.0.

In this chapter, we describe the Trusted Computing Base (TCB) of the PRA-TLS system in Section \ref{tcb}, the capabilities required of the TEE in Section \ref{tee}, and the capabilities required of the Attester Daemon in Section \ref{daemon}. 
We then present the threat model of the PRA-TLS system in Section \ref{threats}.

\begin{table}[t]
\centering
\caption{The trusted model(e.g., Intel SGX).}
\begin{tabular}{|c|c|} \hline
 Root of Trust & Intel SGX CPU, TPM  \\ \hline
 Trust Anchor & Intel SGX: Intel SGX Attestation Root CA \\ \hline
 \multirow{2}{*}{Attestation TCB} & Firmware, Boot loader, Linux Kernel,\\ & System Daemons, and Attester Daemon\\ \hline
 \end{tabular}
 \label{trust_model}
\end{table}

\begin{table*}[t]
\centering
\caption{Functions implemented by existing TEEs.}
\begin{tabular}{|c|c|c|} \hline
  & TEE sealing key & Quote Generation  \\ \hline
 Intel SGX   & \texttt{sgx\_get\_key()} & \texttt{report()}, \texttt{quote()}  \\ \hline
 Arm TrustZone OP-TEE  & PTA\_SYSTEM\_DERIVE\_TA\_UNIQUE\_KEY & OP-TEE Attestation PTA  \\ \hline
 RISC-V Keystone  & \texttt{sbi\_get\_sealing\_key()} & \texttt{sbi\_sm\_attest\_enclave()}  \\ \hline
 \end{tabular}
 \label{system_mapping}
\end{table*}

\begin{table}[t]
\centering
\caption{Usage of PCR in PRA-TLS.}
\begin{tabular}{|c|c|} \hline
 0-7 & Pre-OS boot measurements, including firmware and the boot loader \\ \hline
 8 & Kernel command Line measurements \\ \hline
 9 & Kernel image and Initrd measurements \\ \hline
 10 & Measurements by Linux IMA\\ \hline
 11 & Attester Daemon measurement\\ \hline
 \end{tabular}
 \label{pcr}
\end{table}

\subsection{TCB}
\label{tcb}
We illustrate an example trust model for the PRA-TLS system using Intel SGX as the TEE architecture in \tabref{trust_model}.
The PRA-TLS system uses the TPM as the root of trust to store, protect, and report integrity measurements collected at boot time and during client application execution.
The system also uses the TEE hardware as the root of trust for executing enclaves.
At boot time, Secure Boot verifies the signatures of binaries along the boot chain, including the bootloader and Linux kernel, to ensure that only authorized binaries are executed.
We assume that an attacker cannot modify the Linux kernel after boot and that its integrity remains intact during runtime. Accordingly, we include the Linux kernel in the TCB.

The system also uses the Linux IMA to measure critical software components at boot time, including the Attester Daemon and its dependent libraries.
The trustworthiness of the attestation functionality, including the Attester Daemon, depends on these software components as well as the firmware and bootloader that load and measure them.
We further assume that Linux kernel security mechanisms prevent an attacker from modifying the code or execution state of the Attester Daemon after boot.
Accordingly, we assume that the Attester Daemon remains unmodified and maintains its integrity throughout runtime.

Finally, achieving these guarantees requires ensuring the authenticity and integrity of the Attester Daemon.
The Attester Daemon must access TPM functions to generate attestation evidence. 
To ensure the authenticity of this evidence, the system allows only the Attester Daemon to access the TPM functions required by PRA-TLS, including the use of the AK and the generation of quotes.
To enforce this restriction, the system uses Security-Enhanced Linux (SELinux) \cite{selinux}, a mandatory access control mechanism that controls access based on process domains and resource types, to restrict access to TPM functions to the Attester Daemon.
However, the security of this access control mechanism depends on the authenticity and integrity of the enforced SELinux policy.
Therefore, at boot time, the system uses Linux IMA to measure the SELinux policy files and verifies that they match the expected policy.
The proposed system treats the SELinux policy as part of the Attestation TCB.
We further assume that, after boot, the Linux kernel included in the Attestation TCB continues to enforce SELinux mandatory access control and prevents an attacker from modifying or disabling the SELinux policy.

Accordingly, the TCB of the proposed system consists of the firmware, bootloader, Linux kernel, and Attester Daemon.
We refer to this set of trusted components as the \textbf{Attestation TCB}.
We consider attacks that exploit vulnerabilities in the software stack to compromise the Attestation TCB itself to be out of scope.

\subsection{Required Capabilities of TEE}
\label{tee}
The PRA-TLS system uses two types of enclave code: a TEE Application invoked by the Client Application and a Daemon-side TEE invoked by the Attester Daemon.
The system requires three capabilities from the TEE: confidentiality and integrity protection, a TEE-specific key, and Quote generation.

First, the enclave must protect its code and data from unauthorized access and modification by external software.

Second, the system requires a key available only within the enclave, derived from the identities of the platform and enclave, so that the Verifier can determine whether the evidence originates from the intended execution environment.
The Daemon-side TEE and the TEE Application derive the same key when they run on the same machine and are signed with the same signing key.
Using this key, the TEE application checks that the evidence was generated in response to the attestation request from the target TEE application and client application. 
This prevents the use of evidence relayed from another machine or execution environment due to a malicious client application.
It's described in detail in Section \ref{attack_model-goal}.

Third, the TEE must also support the Quote generation required for enclave attestation, because the proposed system performs client application attestation by extending RA-TLS.
The TEE application stores attestation information about the client application and its execution environment in an X.509 certificate extension used by RA-TLS, in addition to the conventional TEE application attestation information presented to the Verifier.
Appendix \ref{app_ra_tls} provides further details of the X.509 certificate extension used by RA-TLS.
This design enables the Verifier to validate not only the TEE application itself but also the client application's code and execution environment upon receiving the certificate.

\tabref{system_mapping} summarizes the mechanisms provided by existing TEE architectures that can realize the capabilities required by the PRA-TLS model.
The process-based TEE architectures listed in \tabref{system_mapping} provide these capabilities. Based on these common capabilities, we design a generic client application attestation protocol for process-based TEEs.

\subsection{Required Capabilities of the Attester Daemon}
\label{daemon}
This system verifies the authenticity and integrity of the client application invoking the TEE at both deployment time and during execution.
\tabref{pcr} summarizes the PCR values used by the proposed system.
PCR 11 contains the measurement value obtained by extending the Linux IMA measurements of the Attester Daemon and its dependent libraries.
The TPM extend operation combines the current PCR value with a new measurement and hashes the result; therefore, changing the order of measured binaries changes the final PCR value.
Because the OS does not necessarily load dynamic kernel modules in a deterministic order, PCR 10 and PCR 11 may differ even across systems with the same software stack.
Linux IMA records each measurement in an integrity measurement log.
The verifier can validate the accumulated PCR value by replaying the measurements recorded in this log.
In this paper, we assume that the kernel statically links all required modules and does not load any kernel modules dynamically.

The Attester Daemon generates the following two types of evidence for client application attestation.
\begin{itemize}
    \item \textbf{Boot Attestation Evidence}\\
    Evidence derived from PCR values associated with the Attestation TCB, which allows the verifier to determine whether the client application runs on the expected software stack.
    \item \textbf{Runtime Attestation Evidence}\\
    Evidence derived from a hash of the client application code measured by the Attester Daemon, which allows the verifier to assess the integrity of the clinet application itself.
\end{itemize}

We define \textbf{$PCR_{TCB}$} as the set of PCR values consisting of PCRs 0--10, which accumulate measurements of the boot chain from the firmware through the OS kernel and system daemons, and PCR 11, which accumulates Linux IMA measurements of the Attester Daemon and its dependent libraries at boot time.
The Boot Attestation Evidence is a TPM Quote generated using the AK over the selected PCR values represented by \textbf{$PCR_{TCB}$}, together with a nonce.

The Runtime Attestation Evidence is the hash that includes the client application code hash and the Verifier's nonce, which the Attester Daemon receives when it receives an evidence-generation request.

\subsection{Threat Model}
\label{threats}

We do not trust general user-space software, including the client application.
Although the TEE protection mechanism itself does not rely on the OS kernel for security, PRA-TLS depends on kernel-provided measurement and access-control functions. 
We include the Linux kernel in the Attestation TCB.
We also define the boot chain from the firmware through the Linux kernel and the Attester Daemon as the Attestation TCB and assume that these components remain trustworthy.

Following the Dolev-Yao adversary model \cite{dolev-yao}, we assume that an attacker can eavesdrop on, block, modify, and replay messages exchanged between the TEE application and the client application, as well as between the TEE application and the Attester Daemon.
Furthermore, as discussed in Section \ref{motivation}, we consider the boundary between the TEE application and the client application to be an attack surface that a malicious client application can exploit.
When multiple machines have the same Attestation TCB configuration, a malicious client application may relay an attestation request to another legitimate machine and use the evidence generated there as if it were its own. 
To prevent this relay attack, it's necessary to check that the Attester Daemon generated the evidence on the same machine as the TEE application that received the original attestation request.
In this paper, the binding between the TEE application and the Attester Daemon is referred to as ``the Execution Environment Binding''.

TPM modules are often physically attached to the platform board.
We consider hardware-based denial-of-service attacks, such as physically removing or resetting the TPM module, to be out of scope.
We also exclude attacks that physically relocate the TPM to another untrusted machine, thereby invalidating the assumed binding between the TPM-generated evidence and the original platform.

\section{Protocol Overview}
\label{proposed_protocol}
PRA-TLS extends RA-TLS, an existing remote attestation protocol for Intel SGX enclaves, to perform remote attestation of the client application.

In Section \ref{preliminaries}, we describe the preparation required to execute PRA-TLS and the system operations performed at startup.
In Section \ref{protocol}, we then present an overview of PRA-TLS.

\subsection{Preliminaries}
\label{preliminaries}

Before performing client application attestation, we provision the Verifier with the Reference Values required for verification.
These Reference Values consist of the measurement used to identify a legitimate TEE application, the expected PCR values for a correctly booted Attestation TCB, and the hash of the legitimate client application code.
During attestation, the Verifier compares the measurements contained in the evidence with these Reference Values to verify that a legitimate TEE application and client application are running on the expected software stack.

At system startup, Secure Boot verifies signed components along the boot chain, from the firmware through the bootloader to the Linux kernel, and prevents unauthorized software from being executed.
About PCR 10 and 11, Linux IMA measures each component during the boot process and extends the resulting measurements into the TPM PCRs.
After the Linux kernel starts, the system uses Linux IMA to measure software components that form the Attestation TCB, including the Attester Daemon, its dependent libraries, and the SELinux policy.
The Administrator stores the resulting \textbf{$PCR_{TCB}$} as a Reference Value representing the expected state of a correctly configured Attestation TCB before the attestation in the Verifier.
It's also assumed that the hash of the legitimate client application code is stored as a Reference Value in the Verifier.

The system also generates an AK within the TPM for signing Quotes.
It assumes that during provisioning, the system has already verified the EK's authenticity using its EK certificate and established that the generated AK belongs to the same TPM.
The Verifier retains the validated AK public key and uses it to verify the signature on the Boot Attestation Evidence during attestation.
The system also configures a TPM policy that restricts use of the AK to the expected \textbf{$PCR_{TCB}$}, thereby allowing Boot Attestation Evidence to be generated only when the Attestation TCB is in the expected state.

\subsection{Client Application Attestation Protocol}
\label{protocol}

\begin{figure}[t]
  \centerline{\includegraphics[width=\linewidth]{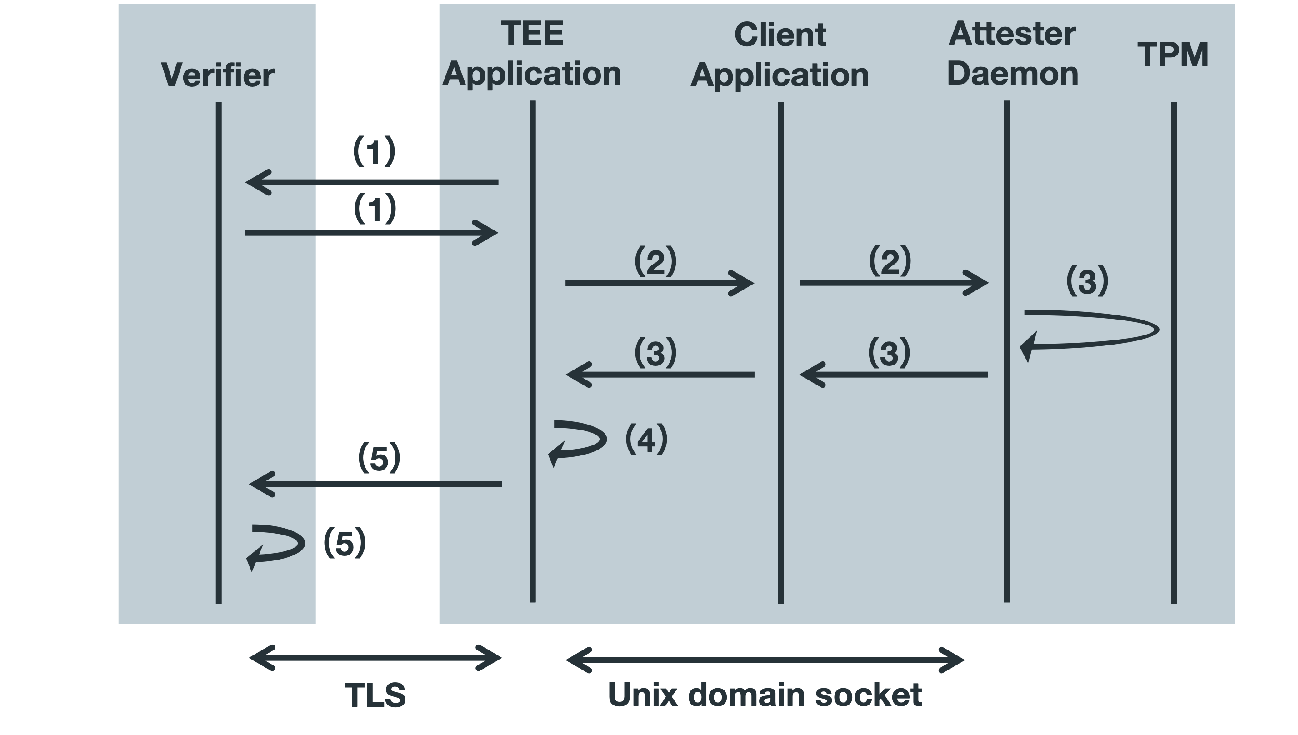}}
 \caption{Attestation protocol of client application.}
 \label{attestation_flow}
\end{figure}

\figref{attestation_flow} illustrates the flow for verifying the client application's attestation.
The protocol involves six entities: the Attester Daemon, which acts as the Attester; the Daemon-side TEE invoked by the Attester Daemon; the Verifier; the client application under verification; the TPM that provides attestation evidence; and the TEE Application that relays and stores the evidence.

The following describes the flow of the proposed client application attestation protocol.

\begin{enumerate}
\renewcommand{\labelenumi}{(\arabic{enumi})}
    \item \textbf{Obtaining a Nonce from the Verifier} \\ 
    Through the TLS handshake, the TEE application sends a ClientHello message to the Verifier and receives a ServerHello message containing a nonce from the Verifier.
    
    \item \textbf{Requesting Evidence} \\
    The TEE application sends the nonce received from the Verifier to the Attester Daemon as an evidence-generation request via the client application over the Unix domain socket.
    
    \item \textbf{Generating and Returning Evidence} \\
    The Attester Daemon requests the TPM to generate a Quote over $PCR_{TCB}$ and the nonce using the AK, which serves as the Boot Attestation Evidence.
    The Attester Daemon also measures the client application code and generates the Runtime Attestation Evidence by hashing the concatenation of the resulting code hash and the nonce.
    The Attester Daemon instructs the Daemon-side TEE to derive a sealing key.
    The Daemon-side TEE encrypts the nonce using the derived sealing key.
    The Attester Daemon then returns the Boot Attestation Evidence, Runtime Attestation Evidence, and encrypted nonce to the requesting TEE application through the client application over the Unix domain socket.
    
    \item \textbf{Verifying the Execution Environment Binding at the TEE application}\\
    The TEE application derives its sealing key and uses it to encrypt the nonce contained in the ServerHello received from the Verifier, thereby obtaining the expected encrypted nonce.
    The TEE application compares the encrypted nonce received from the Attester Daemon with the expected encrypted nonce and verifies that both enclaves derived the same sealing key.
    After successful verification, the TEE application's code embeds the received Boot Attestation Evidence and Runtime Attestation Evidence into X.509 certificate extensions used by RA-TLS and sends the certificate to the Verifier.
    We provide further details on the X.509 certificate extensions used by RA-TLS in Appendix \ref{app_ra_tls}.

    \item \textbf{Attestation of Client Application}\\
    The Verifier uses the AK public key to verify the signature on the Boot Attestation Evidence, thereby confirming its authenticity and integrity, and obtains $PCR_{TCB}$ from the verified evidence.
    The Verifier compares the obtained $PCR_{TCB}$ with the Reference Value of $PCR_{TCB}$ provisioned in advance and verifies that they match.
    The Verifier computes the expected Runtime Attestation Evidence by hashing the reference hash of the legitimate application code together with the nonce and compares the result with the Runtime Attestation Evidence generated by the Attester Daemon.
    Through these verification steps, the Verifier establishes the authenticity and integrity of the client application and its execution environment.
\end{enumerate}

\section{Security Analysis}
In this chapter, we model PRA-TLS and its security properties in the Tamarin prover and formally verify them.

First, Section \ref{tamarin} provides an overview of Tamarin, the formal verification tool used in this study.
Next, in Section \ref{attack_model-goal}, we define the attack model and security requirements for PRA-TLS.
We then describe the Tamarin model of PRA-TLS in Section \ref{modeling} and the modeling of the security properties evaluated in this study in Section \ref{sp-modeling}.
Finally, Section \ref{result} presents the verification results.
Appendix \ref{tamarin_lemma} provides further details of the modeling.
Our model is openly available\footnote{https://github.com/neriyokan/PRA-TLS.git}

\subsection{Formal Verification using Tamarin}
\label{tamarin}

The Tamarin Prover is a state-of-the-art tool for formal verification in symbolic models \cite{symbolic-model}.
The behavior of protocol entities and the adversary model are specified using a multiset rewriting system, while the security properties are expressed as first-order logic formulas.
Tamarin can then automatically verify whether the specified security properties hold in the presence of an unbounded number of concurrently executing protocol instances.
If a security property does not hold, Tamarin provides an attack trace that violates the property as a counterexample.
Tamarin adopts the Dolev-Yao adversary model for modeling attacks on network communication.
In addition, this model assumes perfect encryption, which holds that the attacker cannot recover the plaintext from the ciphertext unless the attacker obtains the key.

A rule describes each protocol step.
A rule represents a state transition that consumes input states or messages and produces subsequent states or messages.
Users can define important protocol events as action facts.
Action facts record events that occur during the transition.
And then they can subsequently be referenced in security properties.
For example, the following rule models an entity \texttt{A} receiving and storing a nonce.
\begin{lstlisting}[firstnumber=1, caption={}, label={}]
rule Example:
    [ Receive(A, nonce) ]
    --[ ReceiveNonce(A, nonce) ]->
    [ Store(A, nonce) ]
\end{lstlisting}
When the rule is applied, the action fact \texttt{ReceiveNonce(A, nonce)} is recorded in the execution trace.
When Tamarin applies rules, it records the resulting sequence of events as a trace.
A trace represents an execution sequence obtained by successive applications of rules.
A lemma expresses a security property as a logical formula over action facts and their temporal ordering in a trace.

Tamarin analyzes protocols with respect to a symbolic model of cryptography.
The properties of the employed cryptographic algorithms are modeled by equations.
In this verification, we use Tamarin’s built-in cryptographic primitives for digital signatures and hashing.
Using these primitives, we model TPM signatures over attestation evidence and cryptographic hashes of the client application code in the proposed protocol.


\subsection{Attack Model and Security Properties}
\label{attack_model-goal}

\begin{table}[t]
\centering
\caption{Security requirements and attack models.}
\begin{tabular}{|l|l|l|} \hline
 Security Goal & Attack model  \\ \hline \hline
 \textbf{G1} Freshness of Evidences & Replay attack  \\ \hline
 \textbf{G2} Integrity of Evidences & Tampering attack \\ \hline
 \textbf{G3} Authenticity of Evidences & Forging attack \\ \hline
 \textbf{G4} Binding between Host Identity and Host code &  Relay attack \\ \hline
 \end{tabular}
 \label{goals}
\end{table}

\tabref{goals} summarizes four security requirements of the proposed protocol and their corresponding attack models.

We consider the following three attack models for PRA-TLS.
\begin{itemize}
    \item \textbf{Replay attack}
    
    An attacker intercepts and stores a previously valid message, then later retransmits it, causing the recipient to accept it as a fresh, legitimate message.

    \item \textbf{Tampering/Forgery attacks} 

    An attacker modifies $PCR_{TCB}$, the client application's code hash, or the nonce in the attestation evidence, or attempts to forge evidence not generated by a legitimate TPM.
    
    
    \item \textbf{Relay attack}

    A malicious client application obtains valid attestation evidence from another execution environment that does not reflect its own runtime state and presents it to the Verifier to evade detection of its compromise.

\end{itemize}

A replay attack can occur when an attacker compromises the client application or can manipulate the communication channel.
In the proposed system, the TEE Application cannot communicate directly with the Attestation Daemon over a Unix domain socket, so it exchanges attestation requests and evidence through the client application.
Therefore, if an attacker compromises the client application or controls the Unix domain socket communication between the client application and the Attester Daemon, the attacker may replay previously generated evidence.
As a result, the attacker may cause the Attester Daemon or Verifier to accept previously generated evidence as a valid response to the current attestation request.
\textbf{G1} defines the security requirement against replay attacks: the protocol must ensure that any evidence accepted by the Verifier was generated in response to the current attestation request.

Evidence tampering and forgery can occur when an attacker compromises the client application or can manipulate the communication channel.
When the Attester Daemon sends evidence to the TEE application or Verifier, an attacker may modify $PCR_{TCB}$ in the evidence or replace it with a forged version.
\textbf{G2} and \textbf{G3} define the security requirements against tampering and forgery attacks and aim to ensure that the protocol detects such attacks.

A relay attack can occur when an attacker compromises the client application.
For example, consider a scenario in which an attacker modifies the host code or one of its dependent libraries, causing a malicious client application to run.
Instead of generating evidence from the execution environment of the malicious client application itself, the attacker relays the attestation request to another machine that has a legitimate Attestation TCB and the expected execution environment.
The attacker then obtains valid evidence generated on the legitimate machine, including valid signatures and measurement values, and presents it to the Verifier as evidence for the malicious client application.
If the Verifier checks only the authenticity and measurement values of the evidence without verifying that the evidence corresponds to the client application and machine actually participating in the attestation session, it may incorrectly accept the malicious client application as legitimate.
\textbf{G4} defines the security requirement against relay attacks.
This requirement ensures that the evidence used for application attestation is generated for the intended target machine and client application, thereby preventing the Verifier from accepting evidence relayed from another machine or generated by a malicious daemon.

\subsection{Modeling the Proposed Protocol}
\label{modeling}

Section \ref{protocol} describes the remote attestation protocol in which an external Verifier attests the client application.
As discussed in Section \ref{threats}, the PRA-TLS model considers the communication channels and the boundary between the TEE and the client application as attack surfaces.
To reflect the threat model in which the client application is untrusted, we model the Unix domain socket communication between the client application and the Attester Daemon, as well as messages exchanged through OCALLs from the TEE application to the client application and ECALLs from the client application to the TEE, as attacker-controlled communication.
We abstract TPM command execution as internal state transitions within the daemon, because SELinux restricts access to the TPM device to the Attester Daemon.

There are some studies that formally verify remote attestation for TEE \cite{ra_tee, ra_tee2, ra_tee3}.
Wilson et al. \cite{ra} formally analyze the authentication properties of protocol agents in remote attestation using Tamarin, based on Lowe's hierarchy of authentication specifications \cite{Lowe}.
Their analysis considers not only conventional two-party remote attestation between a Verifier and a Prover but also three-party protocols involving an Attestation Server, and it additionally requires Verifier authentication.
They also analyze existing commercial remote attestation protocols under a strong adversary model, show that some authentication properties do not hold, and propose a new remote attestation protocol that satisfies the required authentication properties.
Moreover, there is a study that formally verifies the RA-TLS protocol \cite{ra-tls_verified}.
Based on these research results, we omit the modeling of remote attestation for the TEE Application and instead focus on modeling and formally verifying the security of the client application attestation protocol.

When the user is connected to the network, the Verifier and the TEE Application communicate over TLS. 
However, our Tamarin model abstracts away the details of the TLS handshake and cryptographic operations, representing TLS communication as state transitions over a secure communication channel.

We define action facts corresponding to each protocol step and use an \texttt{exists-trace} lemma to verify the existence of an execution trace in which these action facts occur in the order specified by the proposed protocol.
This verification does not establish the security of the proposed protocol itself; rather, it confirms that the modeled protocol admits at least one valid execution trace.
This check also ensures that the security properties described later are not satisfied merely because the model admits no executable protocol trace.

\subsection{Modeling Security Properties}
\label{sp-modeling}

To formally verify that the proposed protocol, as modeled in Section \ref{modeling}, resists replay, tampering/forgery, and relay attacks, we formalize the required security properties.

\subsubsection{Replay Attack}

A replay attack occurs when an attacker stores previously obtained evidence and later retransmits it to the Verifier.
In the proposed protocol, the Attester Daemon generates the Boot Attestation Evidence and Runtime Attestation Evidence using the nonce issued by the Verifier.
During verification, the Verifier checks whether the nonce associated with each piece of evidence matches the nonce that it issued.
In our formal model, we specify two lemmas: \texttt{no\_replay\_attack\_fresh}, which requires that any nonce associated with evidence accepted by the Verifier was previously issued by that Verifier, and \texttt{no\_replay\_attack\_unique}, which requires that evidence associated with the same nonce is not accepted more than once.
These lemmas allow us to verify that replaying previously generated evidence cannot result in successful verification by the Verifier.
Appendix \ref{tamarin_lemma} provides further details of the lemmas that formalize these security properties.
Note that our model does not prevent an attacker from retransmitting messages; rather, we verify that stale, replayed evidence cannot cause the Verifier to accept the attestation.

\subsubsection{Tampering and Forging Attacks}

For evidence integrity, we verify the following properties.
The Attester Daemon generates the Boot Attestation Evidence from $PCR_{TCB}$, which represents the state of the Attestation TCB, and the TPM signs the evidence using the Attestation Key.
The Verifier validates the TPM signature to confirm that an attacker has not modified the evidence after it was generated.
The Verifier further compares $PCR_{TCB}$ in the Boot Attestation Evidence with the corresponding Reference Value to determine whether the Attestation TCB is in the expected state.
Before system deployment, we define the legitimate client application code and provision the Verifier with its hash as a Reference Value.
During attestation, the Verifier compares the hash measured from the executing client application code with the Reference Value to determine whether the host code has been modified after deployment.
This verification allows the Verifier to detect unauthorized modifications to the client application reflected in the Runtime Attestation Evidence.

To ensure evidence authenticity, we define a correspondence property that requires that whenever the Verifier accepts evidence as valid, a legitimate Attester Daemon must have generated the corresponding evidence beforehand.
Together, these properties allow us to verify that the Verifier does not accept tampered or forged evidence as legitimate.
Appendix \ref{tamarin_lemma} provides further details of the lemmas that formalize these security properties.

\subsubsection{Relay Attack}

To detect relay attacks, we verify ``the Execution Environment Binding'', which states that the evidence generated by the Attester Daemon originates from the same machine as the TEE that received the Verifier's attestation request.
When the Daemon-side TEE and the TEE Application invoked by the client application reside on the same machine and use the same key-derivation conditions, they derive the same TEE-specific key.
The Daemon-side TEE therefore encrypts the nonce received from the Verifier using the TEE-specific key, while the Attester Daemon incorporates the nonce into the generated evidence.
The TEE Application then decrypts the encrypted nonce received from the Attester Daemon using its own TEE-specific key.
If the decrypted nonce matches the nonce that the TEE Application sent to the Attester Daemon in the evidence-generation request, the TEE Application confirms that the Daemon-side TEE and the TEE Application derived the same TEE-specific key. 
This, in turn, establishes that the evidence generated by the Attester Daemon originates from the same machine as the TEE that received the Verifier's attestation request.
In contrast, if an attacker relays the attestation request to another legitimate machine, the Daemon-side TEE and the TEE Application derive different TEE-specific keys, causing nonce decryption to fail.
This mechanism detects relay attacks in which valid evidence generated in another execution environment is presented as evidence corresponding to the target TEE.
Appendix \ref{tamarin_lemma} provides further details of the lemmas that formalize these security properties.

\subsubsection{Other Modeling Details}
For the overall correspondence of the proposed protocol, we require that whenever the TEE Application starts processing after attestation, the Verifier must have previously issued the nonce, the Attester Daemon must have generated the corresponding evidence, the TEE Application must have verified it, and the Verifier must have validated the evidence in the prescribed order using the same nonce and corresponding evidence.
This property allows us to verify that the sequence of attestation steps leading to execution by the TEE Application follows the intended ordering and correspondence relationships.

For sealing-key confidentiality, we require that the adversary never obtains any generated sealing key at any point in an execution trace.
This property allows us to verify the confidentiality of the sealing key used exclusively by the Daemon-side TEE and the TEE Application.

\subsection{Results}
\label{result}
The results of the Tamarin verification are as follows.
\begin{itemize}
    \item 
    To confirm that PRA-TLS can execute in the intended order, we used an \texttt{exists-trace} lemma to verify the existence of a valid execution trace.
    Tamarin returned \texttt{verified} for this lemma, confirming that the proposed protocol admits at least one valid execution trace.

    \item 
    We considered replay attacks in which an attacker retransmits a previously obtained Quote or other attestation evidence.
    We defined lemmas requiring that, whenever the Verifier or TEE Application accepts a Quote, the associated nonce must have been issued beforehand and that Quotes associated with the same nonce must not be accepted more than once.
    Tamarin returned \texttt{verified} for these lemmas, showing that replayed Quotes or evidence cannot be successfully accepted by the proposed protocol.

    \item 
    We considered attacks in which an attacker tampers with or forges attestation evidence.
    We defined a correspondence lemma that requires that, whenever the Verifier accepts a tuple consisting of a Quote, $PCR_{TCB}$, a client application's code hash, and a nonce, the Attester Daemon must have previously generated corresponding evidence with the same values.
   The verification confirmed that no execution trace exists in which the Verifier accepts tampered or forged evidence as legitimate.
   We also modeled TPM signature verification, comparison of $PCR_{TCB}$ with its Reference Value, and comparison of the client application's code hash with its Reference Value as preconditions for the Verifier's verification rule.
   Tamarin returned \texttt{verified} for all lemmas concerning evidence integrity and authenticity.

    \item 
    We considered a relay attack in which a malicious client application forwards an attestation request to another legitimate execution environment and presents the valid evidence generated there to the Verifier as its own.
    We defined correspondence properties requiring that, whenever the Verifier accepts evidence, the Daemon-side TEE and the TEE Application must have previously derived the same sealing key, and the TEE Application must have completed verification of the key agreement before the Verifier accepts the evidence.
    Tamarin returned \texttt{verified} for both lemmas.
    These results confirm that a relay attack cannot succeed by forwarding valid evidence generated in a different execution environment and causing the Verifier to accept it as evidence corresponding to the target system.

    \item 
    Tamarin also returned \texttt{verified} for the correspondence property of the overall protocol.
    This result confirms that, whenever the TEE Application starts processing after successful attestation, nonce issuance, evidence generation, verification by the TEE Application, and verification by the Verifier have all occurred beforehand in the prescribed order for the same attestation session.
    For sealing-key confidentiality, Tamarin returned \texttt{verified} for the property requiring that the adversary never obtains any generated sealing key.
    This result confirms the confidentiality of the sealing key used by the Daemon-side TEE and the TEE Application.

\end{itemize}
\section{Implementation}
This section begins with a description of our experimental environments in Section \ref{imp_env}.
Following that, the overview of the PRA-TLS implementation is presented in Section \ref{poc}.
Finally, Section \ref{evaluation} presents an evaluation of our proposed protocol and compares it with existing protocols, including the TLS handshake and RA-TLS.

\subsection{Experimental environment}
\label{imp_env}

\begin{table}[t]
\centering
\caption{Configuration of the servers.}
\begin{tabular}{|c|c|c|} \hline
 \multicolumn{2}{|c|}{Intel SGX Server}  \\ \hline
 CPU &  Intel Xeon Silver 4410Y ( 24 cores, 4 GHz)   \\ \hline
 OS & Ubuntu 22.04.5 LTS  \\ \hline
 Memory &  123 GiB  \\ \hline \hline
 SGX SDK & Intel SGX SDK for Linux 2.26  \\ \hline
 Library & openssl-3.1.8 \\ \hline 
 SGX Library & intel-sgx-ssl (openssl-3.1.8) \\ \hline 
 \end{tabular}
 \label{machine}
\end{table}

\tabref{machine} shows the configuration of the experimental equipment and software used in this experiment.
We use the Intel(R) SGX SDK for Linux \cite{sgxsdk} and the Intel(R) SGX Driver \cite{sgxdriver}, and we perform measurements on two machines.
To minimize the impact of CPU frequency variations during performance measurements, we set the CPU frequency scaling governor for all CPU cores to ``performance'' and pinned the measured process to a specific CPU core.

\subsection{Proof-of-concept Implementation}
\label{poc}

To evaluate the feasibility of the proposed approach, we implement a proof-of-concept (PoC) prototype of PRA-TLS.
We use the OpenSSL library for TLS communication without modifying OpenSSL itself.
In our implementation, the Verifier runs on the server side, while the TEE Application and Client Application, etc., under attestation run on the client-side machine.
During TLS connection establishment, the server-side Verifier requests a client certificate and performs the PRA-TLS attestation procedure on the received certificate.
On the client side, the TEE application receives the TLS server random value, also referred to as the server nonce, together with the server certificate. It verifies the server certificate, generates a PRA-TLS client certificate bound to the server nonce, and sends the client certificate to the Verifier.
This implementation integrates PRA-TLS certificate generation and attestation of both the TEE Application and Client Application into the TLS handshake without modifying the existing OpenSSL handshake implementation.

\subsection{Evaluation}
\label{evaluation}

\begin{table}[t]
    \centering
     \caption{The median and mean latency for each case.}
    \begin{tabular}{|c|c|c|} \hline
          [ms]  &  median & mean \\ \hline
    PRA-TLS     &  1301  & 1257 \\ \hline
TLS handshake   &  2.758 & 2.729\\ \hline
    \end{tabular}
    \label{median_mean}
\end{table}

We measure the handshake latency for two cases: a standard TLS handshake between the Server (Verifier) and Client (TEE application), and a PRA-TLS handshake that additionally includes evidence generation to attest to the client application on the Client side.
By comparing these measurements, we quantify the overhead introduced by evidence generation for attestation.

The latency is measured on the TEE application by recording the execution time immediately before and after the \texttt{SSL\_connection} function. 
The standard TLS 1.3 handshake, the PRA-TLS measurement includes the following processing steps:
\begin{itemize}
    \item The TEE application sends an evidence-generation request to the Attester Daemon via the Client application using an OCALL and a Unix domain socket.
    \item The Attester Daemon generates the evidence, including application mesurement, a TPM Quote, and TPM signature, and returns the generated evidence to the TEE application via the Client application using a Unix domain socket and an ECALL.
    \item The TEE application generates the client certificate.
    \item The TEE application creates an SGX quote for enclave attestation.
\end{itemize}

As shown in \tabref{median_mean}, the median latency of PRA-TLS was 1301 ms, and that of the TLS handshake was 2.758 ms.
This increase corresponds to the overhead introduced by the additional processing in PRA-TLS, including evidence generation for client application attestation and client certificate generation.
In this evaluation, the client generates a static certificate in the TLS handshake, whereas PRA-TLS dynamically creates one.
In addition, the TLS handshake doesn't include the time it takes the client to create a key pair for the certificate.
We measured the time required to generate a 3072-bit RSA key pair when dynamically generating a client certificate.
The key-pair generation took 171.1 ms, while the entire certificate generation process took 172.6 ms. 
This result shows that RSA key-pair generation accounts for most of the time spent generating certificates.

These results demonstrate the performance characteristics of the proposed method and provide a basis for evaluating its applicability to practical systems.

\section{Conclusion and Future work}
In this study, we have proposed PRA-TLS, an extension of RA-TLS, an existing remote attestation protocol for Intel SGX, to attest not only the enclave but also the client application that invokes it and its execution environment.
We defined the attack model and security requirements for PRA-TLS, modeled the protocol using the Tamarin Prover, and formally verified that PRA-TLS satisfies these requirements.
Furthermore, we implemented a proof-of-concept (PoC) prototype of PRA-TLS on Intel SGX and compared its performance against a standard TLS handshake.

As future work, we plan to investigate mechanisms for storing attestation evidence to support offline environments.
In particular, we will investigate mechanisms that maintain attestation continuity across reboots and crashes while preventing the unauthorized reuse of stale evidence or state.
In addition, in the new model, the client certificate is signed with the key created during the TLS handshake, so we plan to use a TPM key pair to sign the certificate.

\vspace{1em}

\begin{acknowledgment}
This work was supported by JST K Program Grant Number JPMJKP24U4, Japan.
We would also like to express our sincere gratitude to Professor Atsushi Igarashi of Kyoto University for his valuable advice on this research.
\end{acknowledgment}

\bibliography{ref}
\bibliographystyle{ipsjunsrt-e}

\appendix

%

\section{Security Lemmas in Tamarin Model}
\label{tamarin_lemma}

Tamarin lemmas can be specified using two types of trace quantification: \texttt{all-traces} and \texttt{exists-trace}.
Tamarin applies \texttt{all-traces} by default and verifies whether the specified formula holds for every possible execution trace.

For example, \texttt{L\_no\_replay\_attack\_fresh} requires that whenever the Verifier verifies evidence associated with a nonce, the Verifier must have issued that nonce beforehand.
Specifically, if \texttt{VerifierVerified(quote, PCR\_tcb, host\_code\_hash, nonce, ak)} occurs at time point \texttt{\#i}, then there must exist an earlier time point \texttt{\#j} at which \texttt{VerifierIssuedNonce(sid, nonce)} occurs for the same nonce, such that \texttt{\#j < \#i}.
If Tamarin returns \texttt{verified}, the property holds for all execution traces admitted by the model.

In these lemmas, host\_code\_hash indicates the hash of the client application and the nonce.

\texttt{VerifierIssuedNonce(sid, nonce)} is an action fact that occurs when the Verifier performs a TLS handshake with the enclave and generates a fresh nonce for the session identified by \texttt{sid}.
\texttt{VerifierVerified(quote, PCR\_tcb, host\_code\_hash, nonce, ak)} occurs when the Verifier receives the evidence and successfully verifies the TPM signature, compares $PCR_{TCB}$ with its Reference Value, and compares the host\_code\_hash with its Reference Value.
Accordingly, the replay-attack lemmas verify that, whenever the Verifier accepts evidence, the corresponding nonce must have been issued by the Verifier beforehand and that evidence associated with the same nonce cannot be accepted more than once.
The following lemmas formalize the security properties against replay attacks.
\begin{lstlisting}[firstnumber=1, caption={Lemma for replay attack}, label={}]
lemma L_no_replay_attack_fresh:
    "All quote PCR_tcb host_code_hash nonce ak #i.
        VerifierVerified(quote, PCR_tcb, host_code_hash, nonce, ak) @ #i
        ==>
        Ex sid #j.
            VerifierIssuedNonce(sid, nonce) @ #j
        &   #j < #i
    "

lemma L_no_replay_attack_unique:
    "All quote1 quote2 PCR_tcb1 PCR_tcb2 host_code_hash1 host_code_hash2 nonce ak #i1 #i2.
        VerifierVerified(quote1, PCR_tcb1, host_code_hash1, nonce, ak) @ #i1
    &   VerifierVerified(quote2, PCR_tcb2, host_code_hash2, nonce, ak) @ #i2
        =>
            #i1 = #i2
    "
\end{lstlisting}

\texttt{CreateEvidences(quote, PCR\_tcb, host\_code\_hash, nonce)} is an action fact that occurs when the Attester Daemon receives an attestation request, obtains $PCR_{TCB}$ and a measurement of the executing host code, and generates the corresponding attestation evidence.
The lemma for tampering and forgery requires that, whenever the Verifier accepts a quote, $PCR_{TCB}$, a host\_code\_hash, and a nonce, the Attester Daemon must have previously generated evidence containing the same values.
This property ensures that the Verifier cannot accept tampered or forged evidence as legitimate.
The following lemma formalizes the security property against evidence tampering and forgery.
\begin{lstlisting}[firstnumber=1, caption={Lemma for tampering and forgery}, label={}]
lemma L_integrity:
    "All quote ak PCR_tcb host_code_hash nonce #i.
        VerifierVerified(quote, PCR_tcb, host_code_hash, nonce, ak) @ #i
         ==>
        Ex #j.
            CreateEvidences(quote, PCR_tcb, host_code_hash, nonce) @ #j
        &  #j < #i
    "
\end{lstlisting}

For relay attack verification, we define \texttt{DaemonSealkey(nonce, sealkey)} and \texttt{EnclaveSealkey(nonce, sealkey)} as action facts that occur when the Daemon-side TEE and the enclave, respectively, obtain the sealing key associated with the attestation request.
\texttt{EnclaveVerified(sid, nonce, ak)} occurs when the enclave verifies the encrypted nonce received from the Attester Daemon using its own sealing key.
\texttt{L\_relay1} verifies that both TEEs use the same sealing key for the same nonce, while \texttt{L\_relay2} verifies that the enclave completes this verification before the Verifier accepts the evidence.
The following lemmas formalize the security properties against relay attack.
\begin{lstlisting}[firstnumber=1, caption={Lemma for relay attack}, label={}]
lemma L_relay_attack1:
    "All quote PCR_tcb host_code_hash nonce ak #i.
        VerifierVerified(quote, PCR_tcb, host_code_hash, nonce, ak)@ #i 
        ==>
        Ex sealkey #j #k.
            DaemonSealkey(nonce, sealkey) @ #j
        &   EnclaveSealkey(nonce, sealkey) @ #k 
        &   #j < #k
        &   #k < #i
    "

lemma L_relay_attack2:
    "All quote PCR_tcb host_code_hash nonce ak #i.
        VerifierVerified(quote, PCR_tcb, host_code_hash, nonce, ak)@ #i
        ==>
        Ex sid #j.
            EnclaveVerified(sid, nonce, ak) @ #j
        &   #j < #i
    "
\end{lstlisting}

\section{PRA-TLS X.509 Certificate Extension}
\label{app_ra_tls}

In Intel SGX RA-TLS, the Verifier validates the enclave according to the Intel SGX ECDSA/DCAP attestation procedure by verifying the Quote, the trust chain based on the Provisioning Certification Key (PCK) certificate, the Certificate Revocation List (CRL), and the TCB information.
The Intel SGX RA-TLS certificate contains the \texttt{sgx-quote} required for enclave attestation and a claims buffer that includes the hash of the TLS certificate public key.
Because \texttt{sgx-quote} also binds the hash of the TLS certificate public key, the Verifier can detect any modification of the hash.

The following defines the extended RA-TLS certificate format used in PRA-TLS, expressed in the Concise Data Definition Language (CDDL) \cite{cddl}.
The fields added to the Intel RA-TLS certificate are highlighted in red.

\begin{lstlisting}[firstnumber=1, caption={the X.509 certificate used in PRA-TLS}, label={}, escapeinside={(*@}{@*)}]
sgx-tpm2-evidence = #6.60000([
    sgx-quote: bstr,
    claims-buffer: bstr .cbor claims
])
claims = {
    "pubkey-hash": bstr .cbor [
            hash-alg-id: uint,	    ; 1: SHA-256
            pubkey-hash: bstr
     ],
     (*@\textcolor{red}{“tpm2-evidences”:bstr .cbor tpm2-evidences}@*)
     (*@\textcolor{red}{“nonce”: bstr}@*)
}
(*@\textcolor{red}{tpm2-evidences = [}@*)
     (*@\textcolor{red}{tpm2-quote: bstr,}@*)
     (*@\textcolor{red}{sign: bstr,}@*)
     (*@\textcolor{red}{app-hash: bstr,}@*)
(*@\textcolor{red}{]}@*)

\end{lstlisting}

This certificate stores \texttt{tpm2-evidences} and the nonce issued by the Verifier.
The \texttt{tpm2-evidences} structure contains three fields: the \texttt{quote} generated by the TPM 2.0, which includes the Boot Attestation Evidence and Runtime Attestation Evidence; the signature on the \texttt{quote} (\texttt{sign}); the hash of the client application code (\texttt{app-hash}).
The \texttt{tpm2-evidences} field corresponds to \texttt{attestationData} in the \texttt{TPM2B\_ATTEST} structure returned by the \texttt{TPM2\_Quote} API and contains a marshaled byte representation of the \texttt{TPMS\_ATTEST} structure.
When the Attester Daemon invokes the \texttt{TPM2\_Quote} API, it specifies the Runtime Attestation Evidence (\texttt{SHA-256(nonce||app-hash)}) as \texttt{qualifyingData}, which the TPM incorporates into the \texttt{extraData} field of \texttt{TPMS\_ATTEST}.
The \texttt{qualifyingData} is stored in the \texttt{extraData} field of the \texttt{TPMS\_ATTEST} structure and is signed by the Quote signature.
During verification, the Verifier unmarshals \texttt{tpm2-evidences}, extracts \texttt{extraData}, and verifies that it matches the expected value \texttt{SHA-256(nonce || app-hash)}.
The \texttt{sign} field contains the marshaled byte representation of the \texttt{TPMT\_SIGNATURE} structure returned by \texttt{TPM2\_Quote}.
In the CDDL definition, \texttt{[]} denotes a CBOR array, whereas \texttt{\{\}} denotes a CBOR map.
The notation \texttt{bstr.cbor\{\}} or \texttt{bstr.cbor[]} indicates a byte string containing the serialized representation of a CBOR map or array, respectively.

\end{document}